\documentclass[12pt,twoside]{article}
\usepackage{fleqn,espcrc1}

\usepackage{epsf}

\newcommand{\AmS}{{\protect\the\textfont2
  A\kern-.1667em\lower.5ex\hbox{M}\kern-.125emS}}
\newcommand{\MeV}{\mbox{MeV}}

\title{Dynamics and Sizes of the Fireball at
Freeze-out\thanks{Work supported by DAAD, BMBF, GSI, and DFG.}}

\author{Boris Tom\'a\v sik\address{Institut f\"ur Theoretische Physik,
	Universit\"at Regensburg, D-93040 Regensburg}%
	\thanks{Address after Oct. 1, 1999: Physics Department, 
	University of Virginia, Charlottesville, VA 22901,
	USA},
	Urs Achim Wiedemann\address{Physics Department, 
	Columbia University, New York, NY 10027.},
	and
	Ulrich Heinz\address{Theoretical Physics Division, CERN, CH-1211 
        Geneva 23}\thanks{On leave of absence from Institut f\"ur 
        Theoretische Physik, Universit\"at Regensburg.}}

\begin{document}

\maketitle

%%%%%%%%%%%%%%% preprint number in upper right corner %%%%%%%%%%%%%%%%%
% Please comment for the version to be submitted to the proceedings
% (will not change vertical position of title)
%%%%%%%%%%%%%%%%%%%%%%%%%%%%%%%%%%%%%%%%%%%%%%%%%%%%%%%%%%%%%%%%%%%%%%%
\begin{flushright}
\vspace*{-7.2cm}
{\footnotesize{\sffamily CERN-TH/99-216\\TPR-99-11\\nucl-th/9907075}}
\end{flushright}
\vspace*{5.2cm}
%%%%%%%%%%%%%%%%%%%%%%%%%%%%%%%%%%%%%%%%%%%%%%%%%%%%%%%%%%%%%%%%%%%%%%%

\begin{abstract}
Analyzing the $m_\perp$-spectrum and two-particle correlations of 
negative hadrons from 158~$A$GeV/$c$ Pb+Pb collisions at slightly 
forward rapidities we find a (thermal) freeze-out temperature of 
about 100~MeV and transverse flow with $\bar v_\perp \approx 0.55c$. 
The $M_\perp$-dependence of the correlation radii prefers a box-like 
transverse density profile over a Gaussian. From an analysis of the pion
phase-space density we find $\mu_\pi \approx 60\, \MeV$ at thermal 
freeze-out.
\end{abstract}

%%%%%%%%%%%%%%%%%%%%%%%%%%%%%%%%%%%%%%%%%%%%%%%%%%%%%%%%%%%%%%%%%%%%%%%%%%%

\section{MOTIVATION}
\label{motiv}

In relativistic heavy-ion collisions the momenta of the escaping hadrons 
are fixed at the point of ``thermal freeze-out''. They carry information 
about the final state of the fireball which one can use to reconstruct 
this state and thus obtain boundary conditions for back-extrapolation 
into earlier stages of its evolution. In this way one can examine the 
question whether or not a quark-gluon plasma was created.

Assuming thermalization and collective transverse flow at freeze-out,
these two features can be extracted separately by combining
single-particle $m_\perp$-spectra and two-particle Bose-Einstein 
correlations \cite{CL96a,CSH95,CNH95,data}. Particles of non-zero 
transverse momentum are emitted from a fireball region which moves 
outwards transversely; in the local comoving frame their transverse 
momentum is lower, resulting in an enhanced Boltzmann factor. 
Transverse flow thus flattens the $m_\perp$-spectrum. A given 
spectral slope can, however, be reproduced by different combinations 
of flow and temperature, because larger flow can be compensated for 
by lower temperature and vice versa. On the other hand, the sizes of 
the regions emitting particles of given momentum (homogeneity regions, 
measured by correlation radii) are controlled by the expansion 
velocity gradients. The natural velocity measure is given by the 
chaotic thermal motion; increasing $T$ and the transverse flow 
(velocity gradients) together at the proper rate will thus not 
change the resulting correlation radii. In summary, $T$ and 
$v_\perp$ are {\em anticorrelated} by a given spectral slope, but 
{\em correlated} by the measured homogeneity sizes. Exploiting both 
types of correlations their values can be determined unambiguously. 
This method was used for analyzing the data from central Pb+Pb
collisions at 158 $A$GeV/$c$ taken by the NA49 collaboration 
\cite{data}.

%%%%%%%%%%%%%%%%%%%%%%%%%%%%%%%%%%%%%%%%%%%%%%%%%%%%%%%%%%%%%%%%%%%%%%%%%%%%%%%

\section{THE MODEL}
\label{model}

It is impossible to extract from Bose-Einstein correlations the global 
fireball features in a model-independent way. We therefore parameterize
the fireball by an {\em emission function}, i.e. by a model for the 
(Wigner) phase-space density of the particles at freeze-out 
\cite{CL96a,TH98,fp}:
 \begin{eqnarray}
   S(x,K) \, d^4x &=& 
   \frac{m_\perp\cosh(y - \eta)}{(2\pi)^3}\, 
   \exp\left ( - \frac{K\cdot u(x) - \mu}{T} \right )\, 
   \exp\left ( - \frac{(\eta - \eta_0)^2}{2\, (\Delta\eta)^2}\right ) 
 \nonumber \\ 
   && \times G(r)\,
   \frac{d\tau}{\sqrt{2\pi(\Delta\tau)^2}}\, 
   \exp\left ( - \frac{(\tau-\tau_0)^2}{2\, (\Delta\tau)^2} \right)\, 
   \tau \, d\eta\, r\, dr\, d\varphi\, .
 \label{e1}
 \end{eqnarray}
The main assumptions here are the following: (i) Approximate thermal
equilibrium at temperature $T$. (ii) Boost-invariant longitudinal 
flow combined with a tunable transverse expansion profile with a 
transverse flow rapidity given by $\eta_t(r) = \eta_f (r/r_{\rm rms})$; 
this expansion is encoded in the velocity field $u(x)$. (iii) For the 
transverse density profile $G(r)$ two different forms were studied:
 \begin{equation}
 \label{e3}
   \mbox{box-like:}\quad 
    G(r) = \theta(R_B-r)\, , \qquad \quad
   \mbox{Gaussian:}\quad
    G(r) = \exp\left ( - \frac{r^2}{2\, R_G^2} \right )\, .
 \end{equation}
If the parameters $T$, $\eta_f$, and $R_B$ or $R_G$ are independent of 
space-time rapidity $\eta$, the model cannot reproduce the rapidity 
dependence of the data. We therefore concentrate here on a single 
rapidity window ($3.9<Y<4.4$). The time parameters $\tau_0$, $\Delta\tau$ 
were fixed together with the other parameters from the correlation 
radii, whereas the longitudinal width $\Delta\eta=1.3$ of the 
density profile was fixed by adjusting the width of $dN/dy$.

The correlation radii were calculated numerically from the
``model-independent expressions'' \cite{CSH95}. When calculating the
single-particle spectrum resonances were taken into account as in
\cite{WH97res}; since we fitted the $h^-$ spectrum we also added 
negative kaons and antiprotons. For the resonances baryonic and 
strangeness chemical potentials, $\mu_{\rm B}$ and $\mu_{\rm S}$, were 
taken into account as required by baryon number and strangeness 
conservation \cite{fp}.

%%%%%%%%%%%%%%%%%%%%%%%%%%%%%%%%%%%%%%%%%%%%%%%%%%%%%%%%%%%%%%%%%%%%%%%%%%
\section{THE FIT}
\label{fit}
%%%%%%%%%%%%%%%%%%%%%%%%%%%%%%%%%%%%%%%%%%%%%%%%%%%%%%%%%%%%%%%%%%%%%%%%%%

\begin{figure}[t]
\begin{center}
\vspace*{-.3cm}
\epsfxsize=15cm
\centerline{\epsfbox{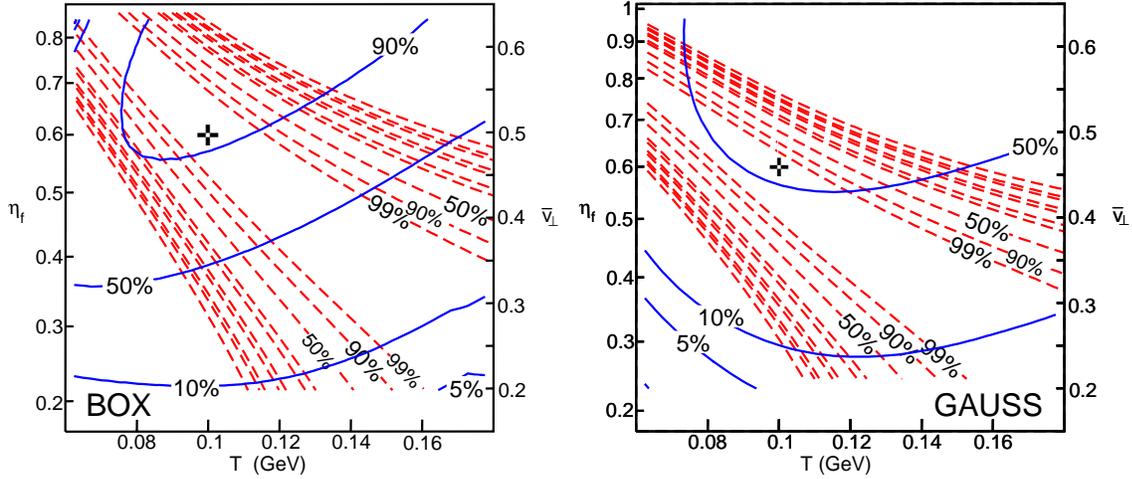}}
\vspace*{-0.5cm}
\caption{$\chi^2$ contours resulting from the fits to the 
single-particle $m_\perp$-spectrum of $h^-$ \cite{jones} (dashed) and
the two-particle correlations (solid). The contours correspond to the
indicated confidence levels. The two plots are obtained with different
transverse density profiles. Crosses show the positions of parameter 
sets used in comparative calculations.}
\label{chi}
\end{center}
\vspace*{-1.5cm}
\end{figure}

Fig.~\ref{chi} shows the $\chi^2$ contours of our fit to the data.
The valley extending from the upper left to the lower right corner 
shows the ambiguity of the extraction of temperature and transverse 
flow from the spectrum. Moreover, the left and right figure show
also a model ambiguity: at fixed temperature different density profiles
lead to slightly different transverse expansion velocities. The 
results of the fit to the correlation radii (width parameters 
in a Gaussian parametrization of the correlation function) indicate 
that a box-like density fits the data better than a Gaussian one, 
although the latter cannot be excluded. Strictly speaking, only 
models with too weak transverse flow seem to be completely wrong. 
With the box-model we find $T\approx 80 - 110\, \MeV$ and 
$\bar v_\perp \approx 0.47c - 0.62c$. The radius of the transverse 
density distribution $R_B = 12.1\, \mbox{fm}$, corresponding to 
$r_{\rm rms}= 8.6$ fm, is about twice as large as the initial
transverse overlap region of the two Pb nuclei, consistent with the
strong transverse flow. 

%%%%%%%%%%%%%%%%%%%%%%%%%%%%% Fig. 1 %%%%%%%%%%%%%%%%%%%%%%%%%%%%%%%%%%%%%%
%%%%%%%%%%%%%%%%%%%%%%%%%% end Fig. 1 %%%%%%%%%%%%%%%%%%%%%%%%%%%%%%%%%%%%%

\begin{figure}
\begin{center}
\vspace*{-0.3cm}
\begin{minipage}{10cm}
\epsfysize=7.4cm
\centerline{\epsfbox{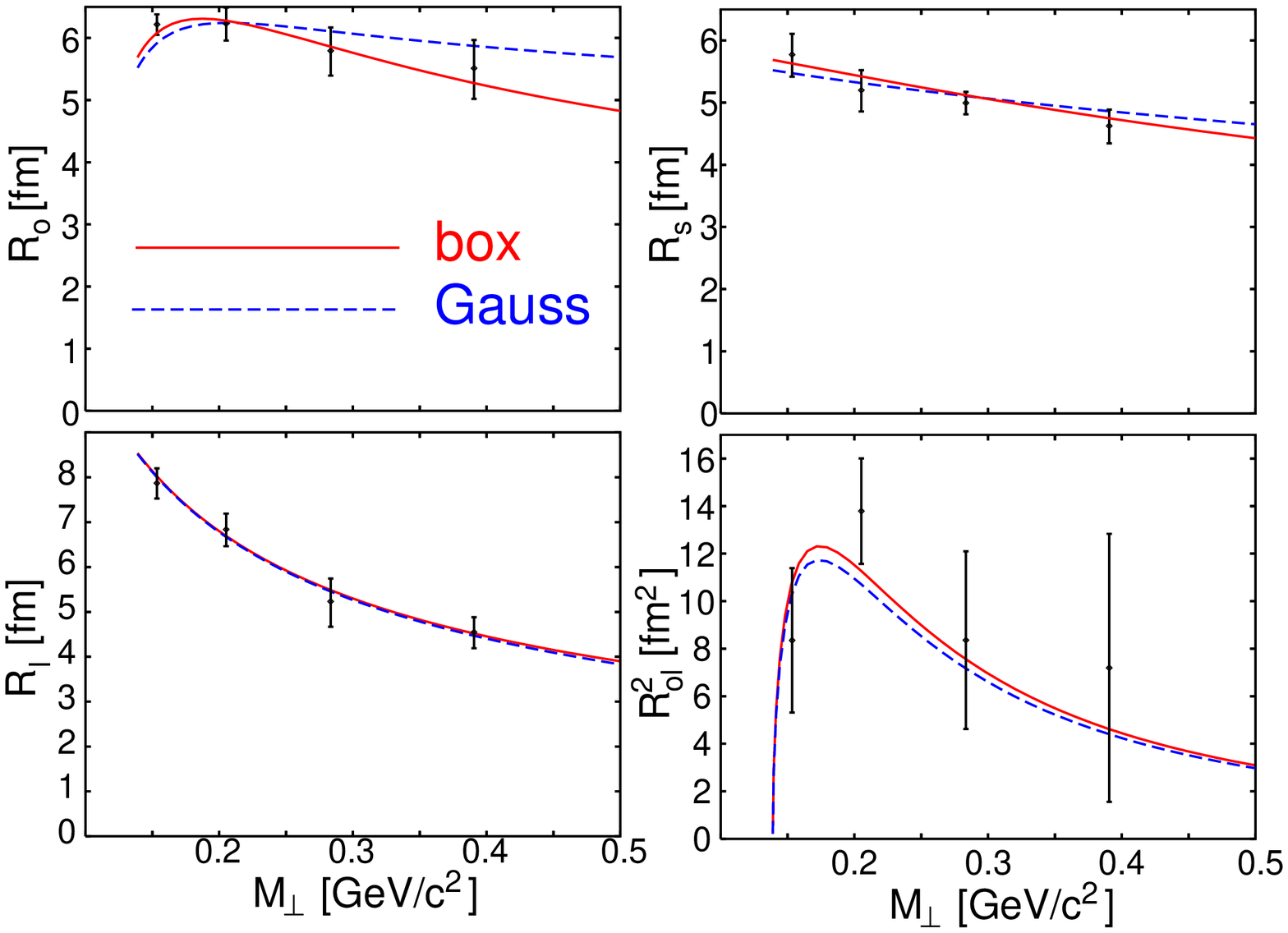}}
\centerline{{(a)}} 
\end{minipage}
\hspace{\fill}
\begin{minipage}{4.5cm}
\epsfysize=7.4cm
\centerline{\epsfbox{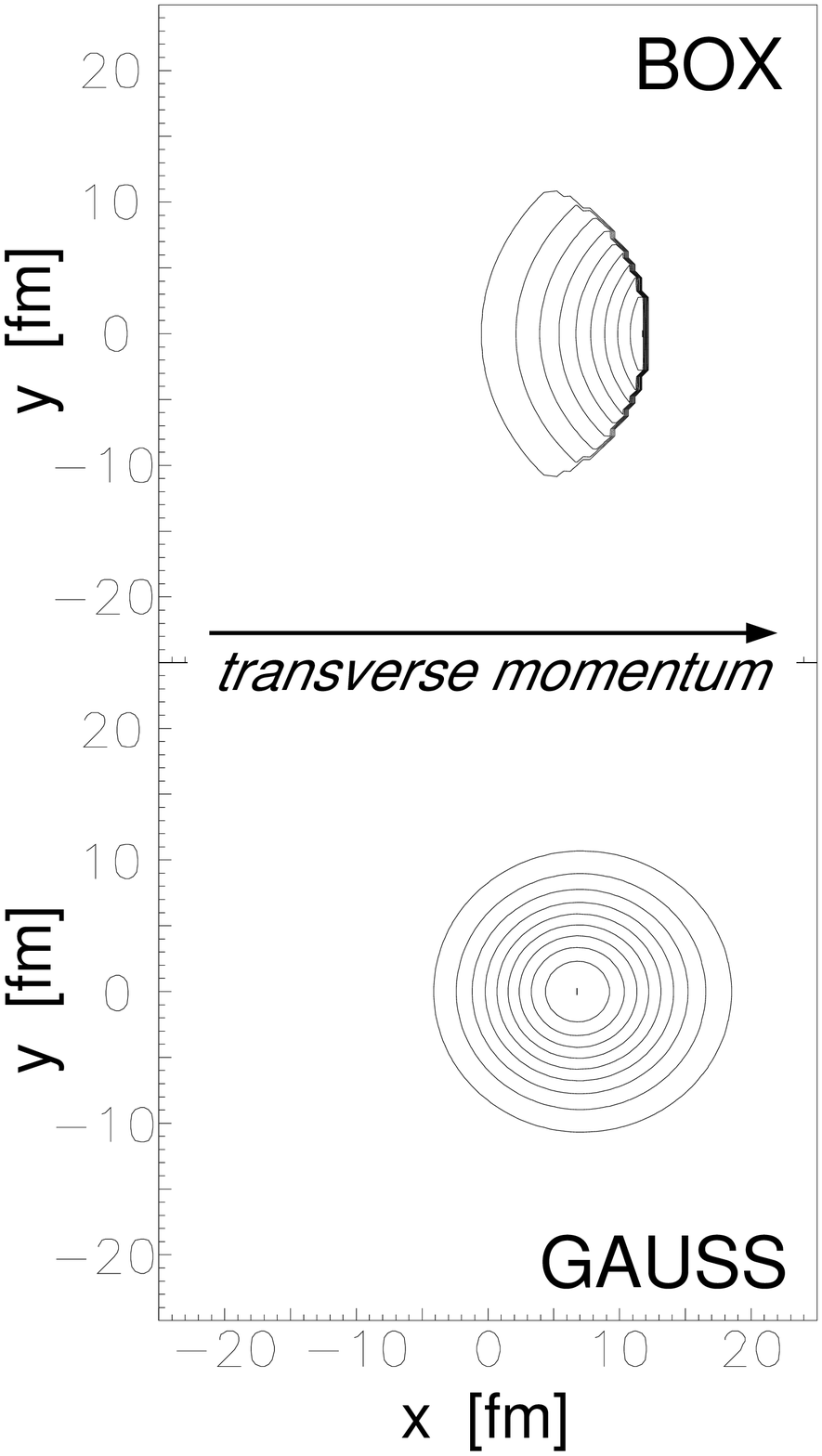}}
\centerline{{(b)}}
\end{minipage}
\vspace*{-.8cm}
\caption{(a) Comparison of the measured correlation radii with the 
two model fits. The temperature ($T$=100 MeV) and transverse flow 
($\bar v_\perp$\,=\,0.5\,$c$ for box, $\bar v_\perp$\,=\,0.46\,$c$ for 
Gauss) of the fits are marked by crosses in Fig.~\ref{chi}. (b) Transverse
cuts of the homogeneity regions in the two models for pions with
$K_\perp = 400$ MeV.}
\label{radcomp}
\end{center}
\vspace*{-1cm}
\end{figure}

Both models fit the $m_\perp$-spectrum well but, as shown in 
Fig.~\ref{radcomp}, they differ in the quality of the fit to the
correlation radii. The $M_\perp$-dependence of transverse correlation 
radii $R_o$ and $R_s$ is better captured by the box-model. This can be
traced back to the fact that the box-model has a rigid radial density 
cutoff; for increasing $K_\perp$ the homogeneity region is thus squeezed 
towards that edge, making it narrower in the direction of $K_\perp$. 
As a result $R_o$ decreases more strongly with $K_\perp$ than for 
the Gaussian model where the homogeneity region expands into the tail
of the density distribution (see Fig.~\ref{radcomp}b). We stress that
it is the generic feature shown in the upper half of Fig.~\ref{radcomp}b 
(and not the box-like transverse density distribution itself) which 
appears to be required by the data.

%%%%%%%%%%%%%%%%%%%%%%%%%%%%%%%% Fig.2 %%%%%%%%%%%%%%%%%%%%%%%%%%%%%%%%%%
%%%%%%%%%%%%%%%%%%%%%%%%%%%%%% end Fig. 2 %%%%%%%%%%%%%%%%%%%%%%%%%%%%%%%%%%

%%%%%%%%%%%%%%%%%%%%%%%%%%%%%%%%%%%%%%%%%%%%%%%%%%%%%%%%%%%%%%%%%%%%%%%%%%%%
\section{MULTIPLICITY}
\label{mult}
%%%%%%%%%%%%%%%%%%%%%%%%%%%%%%%%%%%%%%%%%%%%%%%%%%%%%%%%%%%%%%%%%%%%%%%%%%%%

From the fitted models we can calculate the total pion multiplicity,
assuming chemical equilibrium at thermal freeze-out with 
$\mu_{\rm B} \approx 360\, \MeV$, corresponding to an entropy per baryon of 
about $S/A$=38 \cite{fp,UHqm99}, and $\mu_{\rm S}$ fixed by 
strangeness neutrality. This implies $\mu_\pi$=0 and results in a 
predicted multiplicity which is more than a factor 3 below the measured 
value. By investigating the pion phase-space density \cite{fp} we 
inferred a chemical potential for the direct pions of $\mu_\pi
 \approx 60\,\MeV$ at $T\approx 100\, \MeV$. This implies that at thermal
freeze-out the pions are out of chemical equilibrium. This is expected
if chemical freeze-out happens in equilibrium near the confinement
phase transition at $T_{\rm cr}\approx 170$ MeV \cite{UHqm99,pbm}.

%%%%%%%%%%%%%%%%%%%%%%%%%%%%%%%%%%%%%%%%%%%%%%%%%%%%%%%%%%%%%%%%%%%%%%%%%%%%%
\section{CONCLUSIONS}
\label{conc}
%%%%%%%%%%%%%%%%%%%%%%%%%%%%%%%%%%%%%%%%%%%%%%%%%%%%%%%%%%%%%%%%%%%%%%%%%%%%

A combined analysis of single-particle spectra and two-particle 
correlations avoids the ambiguities and strongly reduces the 
model-dependencies resulting from an analysis of the $m_\perp$-spectrum
of only a single particle species \cite{WA98}. It leads to the unavoidable
conclusion that thermal freeze-out of the fireball happens at a very 
low temperature ($T_{\rm therm}\approx 80-110\, \MeV$) and is accompanied 
by very strong transverse flow ($\bar v_\perp \approx 0.47\,c - 0.62\,c$).  
The $M_\perp$-dependence of the homogeneity lengths can be somewhat 
better reproduced by a box-like than by a Gaussian transverse density 
profile at freeze-out.

At this low temperature the fireball is no longer in chemical equilibrium.
The pion abundance appears to be fixed by chemical freeze-out already
at $T_{\rm chem}\approx T_{\rm cr}\approx 170\, \MeV$ \cite{UHqm99,pbm}. 
We find that at $T_{\rm therm}$=100 MeV this leads to a pion chemical 
potential of $\mu_\pi\approx 60$ MeV \cite{fp,HS}. 

%%%%%%%%%%%%%%%%%%%%%%%%%%%% References %%%%%%%%%%%%%%%%%%%%%%%%%%%%%%%%%%%%


\begin{thebibliography}{99}

\bibitem{CL96a}
  T.\ Cs\"org\H o, B.\ L\"orstad, Nucl. Phys. A590 (1995) 465c; 
  Phys. Rev. C54 (1996) {1390}.

\bibitem{CSH95}
  S. Chapman, P. Scotto, U. Heinz, Heavy Ion Physics 1 (1995) 1.

\bibitem{CNH95}
  S. Chapman. J.R. Nix, U. Heinz, Phys. Rev. C52 (1995) 2694.

\bibitem{data}
  H.\ Appelsh\"auser et al.\ (NA49 Coll.), Eur. Phys. J. C2 (1998) 643;\\
  R.\ Ganz et al. (NA49 Coll.), in Proceedings of the {\it 2nd Catania 
  Relativistic Ion Studies (CRIS'98)}, (eds.\ S.\ Costa et al.), 
  World Scientific, 1999 ({\tt nucl-ex/9808006});\\
  S.\ Sch\"onfelder, PhD thesis, MPI f\"ur Physik, M\"unchen,
  1997;\\
  H.\ Appelsh\"auser, PhD thesis, Universit\"at Frankfurt/Main, 1997.

\bibitem{TH98}
  B.\ Tom\'a\v{s}ik, U. Heinz, Eur. Phys. J. C4 (1998) 327.

\bibitem{fp}
  B. Tom\'a\v{s}ik, U.A. Wiedemann, U. Heinz, CERN-TH/99-215.

\bibitem{WH97res}
  U.A. Wiedemann, U. Heinz, Phys. Rev. C56 (1997) 3265.

\bibitem{jones}
  P.G.\ Jones et al. (NA49 Coll.), Nucl. Phys. A610 (1996) {188c}.

\bibitem{UHqm99}
  U. Heinz, nucl-th/9907060, Nucl. Phys. A (1999), in press.

\bibitem{pbm}
  P. Braun-Munzinger, these proceedings.

\bibitem{WA98}
  H.\ Schlagheck et al. (WA98 Coll.), these proceedings.

\bibitem{HS} 
  C.M. Hung, E. Shuryak, Phys. Rev. C57 (1998) 1891.

\end{thebibliography}
\end{document}